\documentclass[prb,twocolumn,showpacs]{revtex4}
\usepackage{graphicx,bm,amssymb,amsmath}

\newcommand{\Eq}[1]{Eq.~\eqref{#1}}
\newcommand{\eq}[1]{\eqref{#1}}
\newcommand{\Fig}[1]{Fig.~\ref{#1}}

\newcommand{\beq}{\begin{equation}}
\newcommand{\eeq}{\end{equation}}
\newcommand{\beqa}{\begin{eqnarray}}
\newcommand{\eeqa}{\end{eqnarray}}
\newcommand{\Beqa}{\begin{eqnarray*}}
\newcommand{\Eeqa}{\end{eqnarray*}}

\newcommand{\pdag}{{\phantom{\dagger}}}

\newcommand{\hc}{\text{H.c.}}

\newcommand{\bra}[1]{\left\langle{#1}\right\rvert}
\newcommand{\ket}[1]{\left\lvert{#1}\right\rangle}

\newcommand{\PRL}[3]{Phys. Rev. Lett.~\textbf{#1}, #2 (#3)}
\newcommand{\PRB}[3]{Phys. Rev. B~\textbf{#1}, #2 (#3)}

\newcommand{\PR}[3]{Phys. Rev.~\textbf{#1}, #2 (#3)}
\newcommand{\RMP}[3]{Rev. Mod. Phys.~\textbf{#1}, #2 (#3)}

\newcommand{\PhysRep}[3]{Physics Reports~\textbf{#1}, #2 (#3)}

\newcommand{\JETP}[3]{Sov. Phys. JETP~\textbf{#1}, #2 (#3)}
\newcommand{\ZhETF}[3]{Zh. Eksp. Teor. Fiz.~\textbf{#1}, #2 (#3)}
\newcommand{\JPCM}[3]{J. Phys. Condens. Matter~\textbf{#1}, #2 (#3)}
\newcommand{\JPC}[3]{J. Phys. C~\textbf{#1}, #2 (#3)}

\newcommand{\etal}{\textit{et al.}} 

\begin{document}

\title{Kondo temperature of a quantum dot}

\author{Seungjoo Nah} 
\affiliation{
School of Physics, Georgia Institute of Technology, Atlanta, GA 30332, USA}
\author{Michael Pustilnik}
\affiliation{
School of Physics, Georgia Institute of Technology, Atlanta, GA 30332, USA}

\begin{abstract}
We study the dependence of the Kondo temperature on the gate voltage in a strongly blockaded quantum dot with a small single-particle level spacing. We show that the dependence cannot be fitted to that of the Anderson impurity model with the gate voltage-independent level width. The effect originates in high-order tunneling processes, which make a dominant contribution to the exchange amplitude when the gate voltage is tuned away from the middle of the Coulomb blockade valley.
\end{abstract}


\maketitle

\section{Introduction}
\label{Intro}

In a typical transport experiment a nanostructure, such as semiconductor quantum dot,~\cite{blockade,exp_old,exp_new} is connected via tunneling junctions to two massive electrodes. Such devices often exhibit a logarithmic enhancement of the conductance $G$ with lowering the temperature $T$,
\beq
G(T) = G_{0} + \frac{G_K}{\ln^2(T/T_K)}
\label{1}
\eeq
with temperature-independent coefficients $G_{0}$ and $G_K$. This behavior is a manifestation of the well-known many-body phenomenon, the Kondo effect, resulting from the interaction of conduction electrons with an impurity possessing additional degrees of freedom (see Ref.~[\onlinecite{Kondo_review,ABG,Florens}] for a review). The associated energy scale, the Kondo temperature $T_K$, characterizes also the dependencies of the differential conductance on the applied magnetic field and/or source-drain bias, which show a similar logarithmic enhancement.~\cite{footnote} 

The Kondo effect develops when a quantum dot has a non-zero spin in the ground state, which is guaranteed to happen when the dot has an odd number of electrons. The number of electrons $N$ is controlled by the electrostatic potential $V_g$ on the capacitively coupled gate electrode.\cite{blockade,Kondo_review,ABG} In the regime of a strong Coulomb blockade, $N$ is close to an integer at almost any $V_g$ except narrow mixed-valence regions, where adding an electron to the dot is not associated with a large penalty in electrostatic energy. Observable quantities, including $G_0,G_K$, and $T_K$ in \Eq{1}, exhibit a quasi-periodic dependence on the dimensionless parameter $\mathcal N = V_g/\delta V_g$, where $\delta V_g$ is the distance between the mixed-valence regions.~\cite{ABG,Kondo_review} In terms of $\mathcal N$, these regions are narrow intervals of the width $\Delta\ll 1$ about half-integer values of $\mathcal N$.

In this paper we study the dependence of the Kondo temperature of a quantum dot $T_K$ on the gate voltage $\mathcal N$. Surprisingly, the dependence $T_K(\mathcal N)$ has not receive much attention in the literature (see, however, Refs.~[\onlinecite{GHL}] and [\onlinecite{PB}]). In fact, experimental data are often~\cite{exp_old,exp_new} fitted to the expression 
\beq
\frac{T_K(\mathcal N)}{E_C} 
= \sqrt{\frac{\Delta_\mathcal N\Delta}{1-\Delta_\mathcal N}} \exp 
\left[-\,\,\frac{4\Delta_\mathcal N (1-\Delta_\mathcal N)
}{\Delta}\right],
\label{2}
\eeq
originally derived~\cite{Haldane} for the single-level Anderson impurity model.  In \Eq{2}
$E_C$ is the charging energy and $\Delta_\mathcal N$ is the distance in the dimensionless gate voltage to the charge degeneracy point, 
\beq
\Delta_\mathcal N = 1/2 - |\mathcal N - \mathcal N_0|\,,
\label{3}
\eeq
where $\mathcal N_0 $ is an odd integer. \Eq{2} is applicable for $\Delta_\mathcal N$ in the range
\beq
\Delta\ll\Delta_\mathcal N \leq 1/2\,.
\label{3*}
\eeq

Since the Kondo effect is a crossover phenomenon rather than a phase transition, a precise definition of $T_K$ is somewhat arbitrary. In particular, \Eq{2}, as well as \Eq{12} below, is based on the perturbative renormalization group~\cite{Wilson} and defines $T_K$ up to a gate voltage-independent numerical coefficient of the order of unity. The choice of the coefficient does not affect the validity of \Eq{1}, which is applicable in the weak coupling regime of the Kondo effect $T\gg T_K$. The value of the coefficient, however, becomes important in the strong coupling regime $T\ll T_K$. A survey of various definitions of $T_K$ for the Kondo effect with spin$-1/2$ on the dot~\cite{footnote} can be found in Refs.~[\onlinecite{exp_new}] and [\onlinecite{SM}].

Apart from $E_C$, which sets the overall scale, the dependence $T_K(\mathcal N)$ as given by \Eq{2} is completely characterized by a single dimensionless parameter $\Delta$.  Despite its simplicity, \Eq{2} captures the most essential qualitative feature of $T_K(\mathcal N)$: The Kondo temperature has a minimum in the middle of the Coulomb blockade valley $\mathcal N = \mathcal N_0$. At this point $T_K$ is exponentially small, $T_K/E_C \propto\exp(-1/\Delta)$. 

Moreover, although \Eq{2} is inapplicable in the mixed-valence region $\Delta_\mathcal N\lesssim\Delta$ [see \Eq{3*}], it yields an estimate of the energy scale in this regime. Indeed, at $\Delta_\mathcal N\sim \Delta$ the exponential factor in \Eq{2} is of order of unity, while the prefactor is of the order of $ \Delta$, resulting in $T_K\sim E_C\Delta$. For the Anderson model $E_C\Delta$ coincides with the tunneling-induced width of the single-particle energy level in the dot $\Gamma_0$ [see \Eq{16} below]. Since in the mixed-valence regime the Coulomb blockade is partially ``lifted'', the level width $\Gamma_0$ indeed represents the true scale characterizing the low-energy properties of the system.  

Unlike in the Anderson model, the single-particle level spacing in a quantum dot $\delta$ is much smaller than the charging energy $E_C$.~\cite{ABG,Kondo_review} It was shown in Ref.~[\onlinecite{GHL}] that for $\delta\ll E_C$ the dependence $T_K(\mathcal N)$ differs significantly from that prescribed by \Eq{2}. However, this result was obtained in the limit when the contacts between the dot and the leads are almost open.~\cite{ABG,Matveev95} In this limit the dot is in the mixed-valence regime at all values of $\mathcal N$. In other words, the limit considered in Ref.~[\onlinecite{GHL}] corresponds to $\Delta \sim 1$, which is incompatible with the condition \eq{3*} of the validity of \Eq{2}. In this paper we show that, contrary to the widespread belief, \Eq{2} 
is inapplicable to sufficiently large quantum dots in the weak tunneling regime as well, even though \Eq{3*} is satisfied in a wide range of gate voltages.

It should be noted that experiments~[\onlinecite{exp_old,exp_new}] are not performed in the regime of open dot-lead contacts. Indeed, in the open contacts limit the temperature-independent elastic cotunneling~\cite{AN,ABG,Kondo_review} contribution to the conductance $G_0$ [see \Eq{1}] approaches $G(0)/2$; hence, the temperature-dependent Kondo contribution enters \Eq{1} with a vanishingly small coefficient $G_K = G(0) - 2G_0\ll G(0)$ (see Ref.~[\onlinecite{PG}]). Accordingly, in this limit the conductance is almost unaffected by the Kondo effect. On the contrary, in experiments aiming at indisputable realization of the Kondo effect, such as those described in Refs.~[\onlinecite{exp_old}] and [\onlinecite{exp_new}], $G(0)\approx G_K\gg G_0$.  

The rest of the paper is organized as follows: In Section~\ref{model} we describe the model of a quantum dot coupled by tunneling to the conducting leads. In Section~\ref{Kondo} we derive the exchange amplitude of the effective low-energy Kondo model both in the limit $\delta\gg E_C$, corresponding to the Anderson model, and in the limit $\delta\ll E_C$, corresponding to a quantum dot. The results are discussed in Section~\ref{Discussion}.

\section{The model}
\label{model}

We consider a strongly asymmetric configuration, when the conductances of the dot-leads contacts are very different; this simplification does not affect the results. In this case the lead with a weaker coupling to the dot plays the part of a weakly coupled probe, \Eq{1} remains intact,~\cite{Kondo_review} and for the evaluation of the Kondo temperature it is sufficient to consider coupling to a single lead,
\beq
H = H_c + H_d + H_t.
\label{4}
\eeq
Here 
\beq
H_c = \sum_{ks} \xi^\pdag_k c^\dagger_{ks} c^\pdag_{ks}
\label{5}
\eeq
describes electrons in the lead. For a lateral quantum dot system formed by electrostatic depletion of a 2D electron gas at the interface of a semiconductor heterostructure,~\cite{blockade,exp_old} it is sufficient to take into account only a single propagating mode per dot-lead contact,~\cite{Kondo_review,Matveev95,ABG} and $\xi_k$ can be linearized near the Fermi level, which corresponds to a constant density of states $\nu$. 

The second term in \Eq{4} describes an isolated quantum dot.
We consider the simplest model,~\cite{ABG,Kondo_review}
\beq
H_d = \sum_{ns}\epsilon^\pdag_n d^\dagger_{ns}d^\pdag_{ns} 
+ E_C (\hat N - \mathcal N)^2.
\label{6}
\eeq
Here $\hat N = \sum_{ns} d^\dagger_{ns}d^\pdag_{ns}$ is the total number of electrons in the dot and $E_C$ is the charging energy. The single-particle energies $\epsilon_n$ are characterized by a finite level spacing $\delta\ll E_C$. The Fermi level corresponds to $\epsilon_0 = 0$; this level is singly occupied when the number of electrons in the dot $N = \langle \hat N\rangle$ is odd.  

Finally, the last term in \Eq{4} represents the tunneling between the dot and the lead,
\beq
H_t = t_0\sum_{nks}c^\dagger_{ks} d^\pdag_{ns} + \hc
\label{7}
\eeq
A description of the dot-lead contact in terms of the tunneling Hamiltonian \Eq{7} is possible~\cite{ABG,Kondo_review} when the dimensionless (i.e., in units of $2e^2\!/h$) conductance of the contact is small, 
\beq
g = 4\pi \Gamma_0/\delta\ll 1.
\label{8}
\eeq
Accordingly, the tunneling-induced width $\Gamma_0 = \pi\nu t_0^2$ of single-particle energy levels in the dot, the single-particle level spacing $\delta$, and the charging energy $E_C$ form a well-defined hierarchy,
\beq
\Gamma_0\ll\delta\ll E_C.
\label{9}
\eeq

\section{The effective Kondo model}
\label{Kondo}

When the gate voltage is tuned away from the mixed-valence regions, at $\mathcal N$ close to an odd integer $\mathcal N_0$, the dot has an odd number of electrons $N\approx\mathcal N_0$ and its ground state has spin $S = 1/2$. The low-energy excitations of the hamiltonian Eqs. \eq{4}-\eq{7} are then described by the effective Kondo model~\cite{Kondo_review}
\beq
H = H_c + V\rho + J(\textbf{s}\cdot\textbf{S}),
\label{10}
\eeq
where $\rho = \sum_{kk' s} c^\dagger_{ks}c^\pdag_{k's}$ and 
$\textbf{s} = \sum_{kk' ss'} c^\dagger_{ks}(\bm{\sigma}_{ss'}/2)c^\pdag_{k's'}$
represent the local particle and spin densities of conduction electrons, and the spin 
$\textbf{S}$ is the projection of 
$\hat{\textbf{S}} = \sum_{nss'} d^\dagger_{ns}(\bm{\sigma}_{ss'}/2)d^\pdag_{ns'}$
onto the ground-state multiplet of an isolated dot.

The potential scattering term in \Eq{10} is responsible~\cite{PB} for the deviations of $N$ from $\mathcal N_0$, 
\beq
N - \mathcal N_0 \approx -2\nu V.
\label{11}
\eeq
The reduction of the original model \eq{4}-\eq{7} to the effective Kondo model \eq{10} is possible for $|N - \mathcal N_0|\ll 1$, which results in the restriction \eq{3} on the allowed values of $\mathcal N$.

The exchange term in \Eq{10} leads to the Kondo effect characterized by the Kondo temperature~\cite{Wilson}
\beq
T_K \simeq D_0(\nu J)^{1/2} \exp(-1/\nu J).
\label{12}
\eeq
Here $D_0$ is the high-energy cutoff; it corresponds either to the threshold for the intradot excitations $\delta$ or to the energy cost, 
\beq
E_\pm = 2E_C \bigl| \mathcal N - \mathcal N_0 \mp 1/2\bigr|,
\label{13}
\eeq
for adding/removing an electron to/from the dot, whichever is smaller:
\beq
D_0 = \min\{\delta, E_\pm\} = \min\bigl\{\delta, 2E_C\Delta_\mathcal N\bigr\}.
\label{14}
\eeq 

\subsection{Anderson model ($\bm\delta\bm\gg \bm E_{\bm C}$)}
\label{AM}

We discuss first the limit $\delta\gg E_C$, corresponding to the Anderson impurity model.~\cite{Haldane} Although, in view of \Eq{9}, this limit does not correspond to a realistic situation, it leads to the qualitatively correct dependence $T_K(\mathcal N)$, see the discussion above.  At $\delta\gg E_C$ all but $n=0$ energy levels in the dot are either empty or doubly occupied. Projecting these levels out, we write
\[
\hat N \,\to\, n_\uparrow + n_\downarrow + \mathcal N_0 - 1,
\] 
where $n_\uparrow$ and $n_\downarrow$ are the spin-up and spin-down occupations of $n=0$ level. Substitution into \Eq{6} yields, up to a constant,
\beq
H_d 
= 2E_C \bigl[n_\uparrow n_\downarrow 
- (\mathcal N - \mathcal N_0 + 1/2)(n_\uparrow + n_\downarrow)\bigr].
\label{16*}
\eeq
The tunneling \Eq{7} induces transitions to states with $n=0$ level being empty and doubly-occupied. The transitions are virtual and can be accounted for perturbatively. In the second order in the tunneling amplitude $t_0$ one finds~\cite{SW,Haldane} \Eq{10}, with
\beq
J = 2 t_0^2\!\left(\!\frac{1\,}{\,E_+} + \frac{1\,}{\,E_-}\!\right),
\quad
V =  - \,\frac{t_0^2}{2}\!\left(\!\frac{1\,}{\,E_+} - \frac{1\,}{\,E_-}\!\right),
\label{15}
\eeq
and with $E_\pm$ given by \Eq{13}.
Reduction to the Kondo model is justified when $|N - \mathcal N_0 |\ll 1$. With the help of \Eq{11} and the second equation in \eq{15}, this condition translates into \Eq{3} with the width of the mixed-valence region given by
\beq
\Delta = \Delta_0 = \frac{4}{\pi}\frac{\Gamma_0}{\,E_C}\,.
\label{16}
\eeq
Comparison with the first equation in \eq{15} shows that for the Anderson model
\beq
\Delta =  \nu J_0,
\quad
J_0 = \min\bigl\{J(\mathcal N)\bigr\} = J(\mathcal N_0).
\label{19*}
\eeq

Using Eqs. \eq{3} and \eq{16}, the exchange amplitude \eq{15} is written as
\beq
\nu J = \frac{\Delta}{4\Delta_\mathcal N(1-\Delta_\mathcal N)}\,.
\label{17}
\eeq
Equations \eq{12}, \eq{14}, and \eq{17} then yield \Eq{2} above. Accounting for higher order in $t_0$ contributions results in a correction to $J$ in \Eq{15}. The correction is small, $\Delta J_0/J_0 \sim \nu J_0 \ll 1$,~\cite{Haldane, GA} and its effect on the value of of the Kondo temperature $T_K$ is negligible.

\subsection{Quantum dot ($\bm\delta\bm\ll \bm E_{\bm C}$)}
\label{QD}

The Anderson model result for $J_0 = \min\bigl\{J(\mathcal N)\bigr\}$ [see Eqs. \eq{19*}] remains intact even in the limit $\delta\ll E_C$, when the model is no longer applicable. The leading correction now reads~\cite{GA} $\Delta J_0/J_0 \sim \Gamma_0/\delta \ll 1$. Although the correction is still small, it is by a factor $E_C/\delta\gg 1$ larger than that in the Anderson model and results in an increase~\cite{GA} of $T_K$ in the middle of the Coulomb blockade valley by the factor $C$ with $\ln C = (\Delta J/J_0)/\nu J_0 \sim E_C/\delta \gg 1$.

The effect of the higher order in tunneling contributions to the exchange amplitude turns out to be even more dramatic when $\mathcal N$ is tuned away from the middle of the Coulomb blockade valley $\mathcal N = \mathcal N_0$. Indeed, it well known~\cite{GM,Matveev} that when $\mathcal N$ is close to the mixed valence region, say, at $\mathcal N \approx \mathcal N_0 +1/2$,
transitions between the two almost degenerate charge states of the dot result in diverging logarithmic corrections to the tunneling amplitude. The origin of these corrections is again the Kondo effect, with the two charge states playing the part of the impurity spin.~\cite{Matveev,Matveev95}

Following Ref.~[\onlinecite{Matveev}], we project out virtual transitions to the state with $\mathcal N_0-1$ electrons in the dot, associated with the energy cost $E_-\approx 2E_C\gg E_+$. This amounts to the introduction of a high-energy cutoff in Eqs. \eq{4}-\eq{7}: $|\xi_k|,|\epsilon_n| < 2E_C$.  The projected Hamiltonian can be cast in the form of an anisotropic two-channel spin-$1/2$ Kondo model with the physical spin $s$ representing the channel index,~\cite{Matveev}
\beq
H = \sum_{sp\alpha}\varepsilon^\pdag_{p\alpha} 
\psi^\dagger_{sp\alpha}\psi^\pdag_{sp\alpha}
+ I^\pdag_z \tau^z \hat T^\pdag_z + 
\frac{\,I_\perp}{2\,}\left(\tau^+\hat T^\pdag_- + \tau^-\hat T^\pdag_+\right).
\label{18} 
\eeq
Here the ``bare'' (corresponding to the bandwidth $D = 2E_C$) values of the coupling constants are $I_\perp = 2t_0$, $I_z = 0$. In terms of $\ket{\Downarrow}$ and $\ket{\Uparrow}$, representing, respectively, charge states with $\mathcal N_0$ and $\mathcal N_0 +1$ electrons in the dot, the \textit{pseudospin} operators in \Eq{18} are given by
\[
\hat T_z = \frac{1}{2}\bigl(\ket{\Uparrow}\!\bra{\Uparrow} 
- \ket{\Downarrow}\!\bra{\Downarrow}\bigr),
\quad
\hat T_+ = \hat T_-^\dagger = \ket{\Uparrow}\!\bra{\Downarrow}.
\]
The operators $\psi$ in \eq{18} are the relabeled operators $c$ and $d$ of Eqs. \eq{5}-\eq{7},
\[
\psi_{s,p,\alpha = \Uparrow} = d_{n\to p,s}\,,
\quad
\psi_{s,p,\alpha = \Downarrow} = c_{k\to p,s}\,.
\]
Accordingly, the single-particle energies 
$\varepsilon_{p,\alpha} = - \varepsilon_{-p,\alpha}$
are characterized by the pseudospin-dependent density of states
$\nu_\Uparrow = 1/\delta$, $\nu_\Downarrow = \nu$.
Finally, the local pseudospin density is given by
$
\bm{\tau} = \sum_s\sum_{pp'\alpha\alpha'} 
\psi^\dagger_{sp\alpha}\bigl(\widetilde{\bm{\sigma}}_{\alpha\alpha'}/2\bigr)\,\psi^\pdag_{sp'\alpha'},
$
where components of the vector $\widetilde{\bm{\sigma}}$ are the Pauli matrices acting on the pseudospin degree of freedom. 

The scaling equations for the model \eq{18} read
\beq
\frac{\,d\widetilde I_z}{d\zeta} 
= \widetilde I_\perp^{\,2}\bigl(1-\widetilde I_z\bigr), 
\quad
\frac{\,d\widetilde I_\perp}{d\zeta\,} = \widetilde I_\perp\!
\Bigl[\widetilde I_z - \frac{1}{2}\bigl(\widetilde I_z^{\,2} + \widetilde I_\perp^{\,2}\bigr)\Bigr], 
\label{19}
\eeq
where $\zeta = \ln(2E_C/D)$, and 
\[
\widetilde I_z = \frac{1}{2} (\nu_\Uparrow + \nu_\Downarrow) I_z,
\quad
\widetilde I_\perp = (\nu_\Uparrow \nu_\Downarrow)^{1/2} I_\perp
\]
are dimensionless coupling constants. 
Since SU(2) symmetry is broken, renormalization generates also corrections of the type 
$\sum_{pp'\alpha}\psi^\dagger_{sp\alpha}\psi^\pdag_{sp'\alpha} \hat T_z$. These terms lead to small pseudospin-dependent corrections to the density of states,~\cite{PAK} which we neglect. 

Equations \eq{19} yield
\beq
\widetilde I_\perp^2(D) = \gamma^2 + \widetilde I_z^2(D),
\label{20}
\eeq
where 
\[
\gamma = \widetilde I_\perp(0) = \sqrt{4\Gamma_0/\pi\delta}
\] 
and
\beq
\widetilde I_z(D) = \left\lbrace
\begin{array}{lr}
\gamma^2\ln(2E_C/D), & ~~D\gg E_C\sqrt{\Gamma_0/\delta}\,,
\\
\\ 
~\bigl[\ln(D/T_C)\bigr]^{-1}, & ~~D\lesssim E_C\sqrt{\Gamma_0/\delta}\,.
\end{array}
\right.
\label{21}
\eeq
Here 
\beq
T_C = 2E_C \gamma e^{-\pi/2\gamma}
\label{25*}
\eeq
is the energy scale for the charge Kondo effect~\cite{Matveev} and $D\sim E_C\sqrt{\Gamma_0/\delta}$ is the value of the cutoff at which the two contributions to $\widetilde I_\perp$ in \Eq{20} are of the same order of magnitude, $\widetilde I_z(D)\sim\gamma$. 

Equations \eq{20} and \eq{21} are valid in the weak coupling regime $\widetilde I_z(D)\ll 1$ and as long as the bandwidth $D$ exceeds both the single-particle level spacing in the dot $\delta$ and the addition energy $E_+$. (For $\widetilde I_z(D)\ll 1$ the Knight shift-like renormalization $E_+ \to (1-\widetilde I_z)E_+$ can be neglected.) At smaller $D$, 
\beq
D\lesssim D_* = \max\bigl\{\delta,E_+(\mathcal N)\bigr\},
\label{22}
\eeq
the Hamiltonian is given by \Eq{18} with renormalized coupling constants. At $E_+\ll E_-$, it is equivalent to Eqs. \eq{4}-\eq{7} with the high-energy cutoff $D_*$, and with the tunneling amplitude $t_0$ replaced by the gate voltage-dependent $t$, which satisfies
\beq
4\nu t^2(D_*) = \delta\,\widetilde I_\perp^{\,2}(D_*). 
\label{23}
\eeq
[We neglect a weak potential scattering arising from the $z$-component of the exchange in \Eq{18}.]

Further reduction of the bandwidth from $D_*$ down to $D_0$ [see \Eq{14}] can be carried out without regard to the presence of multiple energy levels in the dot as these levels have been already accounted for in the renormalization of the tunneling amplitude. 
Projecting out the dot's excitations with the help of the Schrieffer-Wolf transformation,~\cite{SW,Haldane} we end up with the Kondo model \Eq{10} with the exchange and potential scattering amplitudes 
\beq
\nu J = \frac{2\nu t^2(D_*)}{E_+},
\quad
V =  -\,\frac{\nu t^2(D_*)}{2E_+},
\label{24}
\eeq  
cf. \Eq{15}.
Repeating the arguments that led to \Eq{16} above, we find the renormalized width of the mixed-valence region,
\beq
\Delta = \frac{4}{\pi}\frac{\Gamma}{\,E_C},
\quad
\Gamma = \pi\nu t^2(\delta)\,.
\label{25}
\eeq

Further analysis depends on the parameters of the system. The richest behavior is realized in the limit
\beq
(\delta/E_C)^2 \ll \Gamma_0/\delta \ll 1.
\label{26}
\eeq
In quantum dots formed by electrostatic depletion of 2D electron gas,~\cite{blockade} the left-hand side of \Eq{26} is controlled by the size of the dot $L$, $\delta/E_C\propto 1/L$ for sufficiently large dots,~\cite{Kondo_review,ABG} while the right-hand side is proportional to the conductance of the dot-lead contact [see \Eq{8}]. Experimentally, these quantities are tuned independently of each other,~\cite{Kondo_review,blockade} and the inequalities \eq{8} and \eq{26} can be satisfied simultaneously.

Equation \eq{26} is equivalent to $\ln(\delta/T_C) \ll \sqrt{\delta/\Gamma_0}$, which is compatible with the assumption that $D_*\sim\delta$ belongs to the weak coupling regime of the charge Kondo effect $\ln(\delta/T_C)\gg 1$. The renormalized level width $\Gamma$ introduced in \Eq{25} is then given by 
\beq
\Gamma = \frac{\pi}{4}\frac{\delta}{\ln^2(\delta/T_C)}
\label{27}
\eeq
and satisfies $\Gamma_0\ll\Gamma\ll\delta$. 

Close to, but still well outside the mixed-valence region, the exchange constant \eq{24} takes the form
\beq
\nu J = \frac{\Delta\,}{\,4\Delta_\mathcal N}\,, 
\quad
\Delta\ll\Delta_\mathcal N \ll \delta/E_C,
\label{28}  
\eeq
similar to that in the Anderson model at $\Delta_\mathcal N\ll 1$ [see \Eq{17}]. Note, however, that $\Delta$ in \Eq{28} is much larger than its Anderson model value $\Delta = \Delta_0$ given in \Eq{16}.

Further away from the mixed-valence region Eqs. \eq{20}-\eq{24} give
\beq
\nu J = \frac{\,\Delta_0}{4\Delta_\mathcal N}
\Bigl[1 + (E_C/\delta)\Delta_0\ln^2\!\Delta_\mathcal N\Bigr],
\quad
\sqrt{\Gamma_0/\delta}\ll\Delta_\mathcal N \ll 1.
\label{29}
\eeq 
The gate voltage-dependent logarithmic correction in \Eq{29} remains small for all $\mathcal N$. 

Stretching \Eq{29} beyond its domain of applicability, and taking into account that at $\Delta_\mathcal N = 1/2$ \Eq{29} represents only a half of the exchange amplitude in the effective Kondo model (at this point $E_+\approx E_-$ and transitions between charge states with $\mathcal N_0$ and $\mathcal N_0 -1$ electrons in the dot make an identical contribution to $J$), we recover the result of Ref.~[\onlinecite{GA}], $\Delta J_0/J_0 \sim \Gamma_0/\delta$. This strongly suggests that the corrections considered in Ref.~[\onlinecite{GA}] and the ones studied in the present paper have a common origin and serves as an independent check that the procedure outlined above indeed captures the dominant contributions to the exchange amplitude.

\section{Discussion}
\label{Discussion}

The gate voltage-dependent corrections to the exchange amplitude, and, therefore, to the Kondo temperature $T_K$, originate in the strong renormalization of the level width. Indeed, close to the middle of the Coulomb blockade valley $T_K(\mathcal N)$ is governed by the Anderson model expression \Eq{2} with $\Delta$ given by its bare single-level value $\Delta_0\sim\Gamma_0/E_C$ [see Eqs. \eq{16} and \eq{29}]. However, the width of the mixed-valence region $\Delta\sim\Gamma/E_C\gg\Delta_0$ depends on the much larger renormalized level width $\Gamma\gg \Gamma_0$, see Eqs. \eq{25} and \eq{27}. 

Similar to the Anderson model, the width $\Gamma$ represents the energy scale in the mixed-valence regime, while the corresponding value of $\Delta$ parametrizes the dependence $T_K(\mathcal N)$ close to but still well outside the mixed-valence region [see \Eq{28}]. Unlike in the Anderson model, cf. \Eq{19*}, there is no simple relation between $\Delta$ and $\nu J_0 = \min\bigl\{\nu J(\mathcal N)\bigr\}\approx \Delta_0$; the latter determines the value of $\min\bigl\{T_K(\mathcal N)\bigr\}$.  

\begin{figure}
\includegraphics[width=0.99\columnwidth]{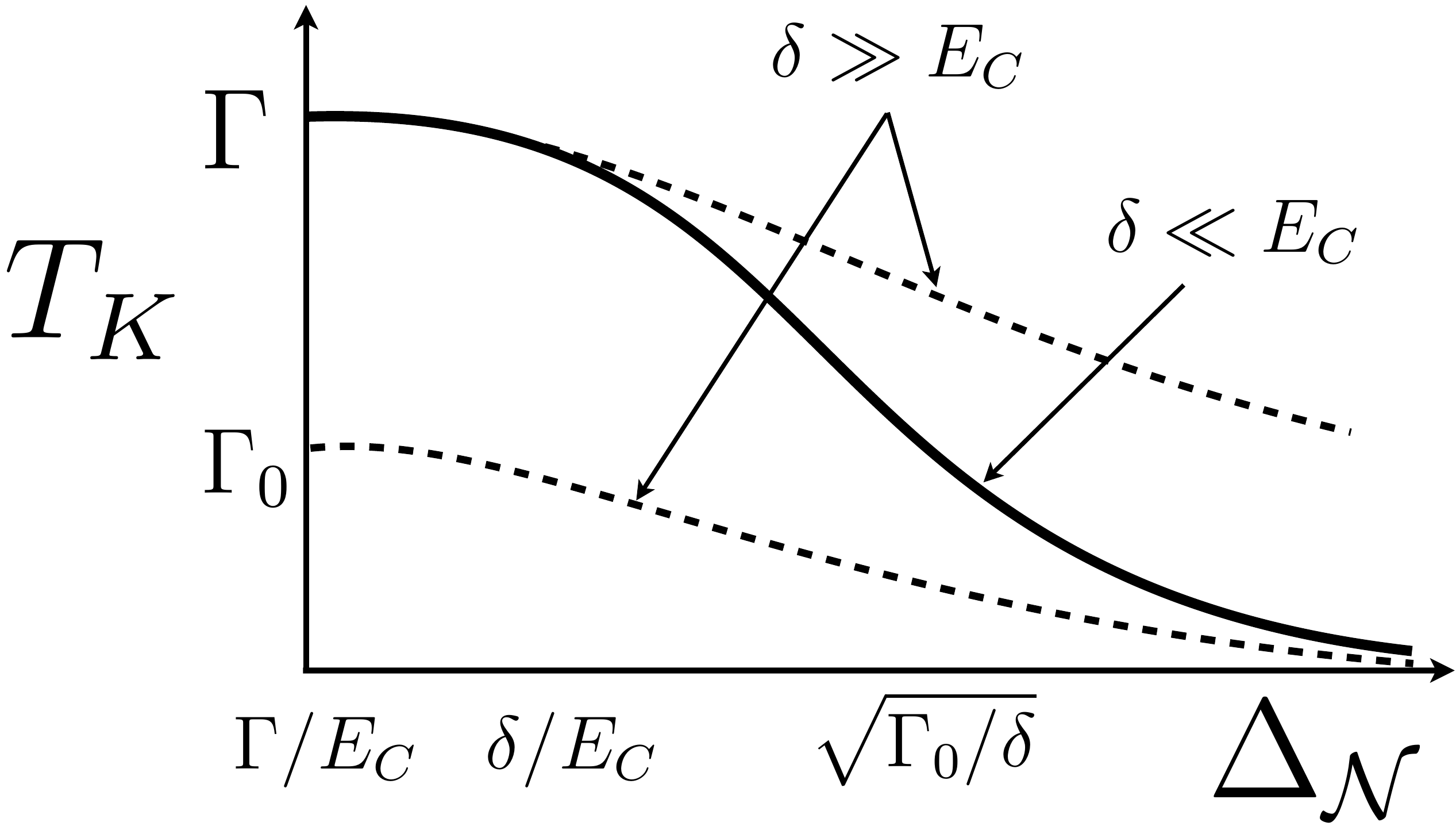}
\caption{
\label{fig1}
Kondo temperature $T_K$ as function of the distance $\Delta_\mathcal N$ to the closest charge degeneracy point. 
The dashed lines correspond to single-level Anderson impurity models with two different values of the tunneling-induced level width, $\Gamma$ and $\Gamma_0$. The solid line represents a quantum dot with a small level spacing $\delta\ll E_C$.
}
\end{figure}

These observations imply that the dependence $T_K(\mathcal N)$ cannot be fitted to the Anderson model's result \Eq{2} with the gate voltage-independent $\Delta$. Instead, $T_K(\mathcal N)$ interpolates smoothly between the corresponding curves for two different Anderson models, as sketched in \Fig{fig1}. This behavior appears to be consistent with the results of the experiments.~\cite{exp_old,exp_new}

In the above derivation we assumed that the tunneling amplitudes $t_n$ are identical for all energy levels in the dot, $t_n = t_0$ [see \Eq{7}]. In large quantum dots with chaotic motion of electrons the amplitudes $t_n$ are not only different, but random and statistically independent of each other.~\cite{ABG,Kondo_review,Alhassid} Accounting for these mesoscopic fluctuations does not affect our results qualitatively. The main difference is that $\Gamma_0$ and $\Gamma$ are now random, although the relation $\Gamma > \Gamma_0$ still holds. Whereas the bare level width $\Gamma_0$ has a broad Porter-Thomas distribution,~\cite{ABG,Kondo_review,Alhassid} the renormalized width $\Gamma$, being a sum of many statistically independent contributions, is expected to have a narrow Gaussian distribution.~\cite{KG} Accordingly, the exchange amplitude $J$ (and, therefore, $T_K$) become less random as the gate voltage approaches the mixed-valence regions.  Note also that for $t_n\neq t_{-n}$ the function $T_K(\mathcal N)$ is no longer symmetric about the middle of the Coulomb blockade valley, 
$T_K(\mathcal N)\neq T_K(2\mathcal N_0 - \mathcal N)$.

Finally, as discussed above, $T_K(\mathcal N)$ in quantum dots with almost open contacts also differs~\cite{GHL} from that in the Anderson model. In this case the conductance of the dot-lead contact is large, $1-g\ll 1$, and 
\[
(\nu J)^{-1}\sim (E_C/\delta)(1-g)\sin^2(\pi\Delta_\mathcal N).
\]
Description of the crossover between the result of Ref.~[\onlinecite{GHL}] and our Eqs.~\eq{24}-\eq{29}, applicable for $g\ll 1$, requires an understanding of an intermediate regime between the strong and weak Coulomb blockade. Despite some progress in this direction,~\cite{intermediate} a detailed theoretical description of this regime remains an open problem.  

To conclude, in this paper we demonstrated that the gate voltage dependence of the Kondo temperature of a quantum dot with a small level spacing is drastically different from that in the Anderson impurity model. 



%

\begin{thebibliography}{99}
\bibitem{footnote}
While \Eq{1} is applicable to all quantum impurity models in the weak coupling regime $T\gtrsim{T_K}$, the behavior of the conductance in the strong coupling regime $T\ll{T_K}$ depends on the type of the Kondo effect realized in a given system, see, e.g.,~[\onlinecite{Kondo_review,Florens}] and references therein.
\end{thebibliography}
\end{document}